\documentclass[10pt]{iopart}
\usepackage{iopams}

\usepackage{graphicx}
\usepackage{booktabs}
\usepackage{subfigure}

\graphicspath{{images/}{plots/}}

\newcommand{\avg}[1]{\ensuremath{\left< #1 \right>}}
\newcommand{\var}[1]{\ensuremath{\mathrm{Var}(#1)}}

\begin{document}

\title[Convex hull of a run-and-tumble particle]{The convex hull
of the run-and-tumble particle in a plane}

\author{Alexander K Hartmann $^1$, Satya N Majumdar$^2$, Hendrik Schawe$^{1,3}$, and Gr\'egory Schehr$^2$}
\address{$^1$ Institut f\"ur Physik, Universit\"at Oldenburg, 26111 Oldenburg, Germany}
\address{$^2$ LPTMS, CNRS, Univ.~Paris-Sud, Universit\'e Paris-Saclay, 91405 Orsay, France}
\address{$^3$ Laboratoire de Physique Th\'{e}orique et Mod\'{e}lisation, UMR-8089 CNRS-Universit\'{e} Cergy Pontoise, France}
\ead{alexander.hartmann@uol.de}
\ead{satya.majumdar@u-psud.fr}
\ead{hendrik.schawe@u-cergy.fr}
\ead{gregory.schehr@u-psud.fr}

\begin{abstract}
    We study the statistical properties of the convex hull of a planar run-and-tumble particle (RTP),
    also known as the ``persistent random walk'', where
    the particle/walker runs ballistically between
    tumble events at which it changes its direction randomly. We consider two different statistical
    ensembles where we either fix (i) the total number of tumblings $n$ or (ii)
    the total duration $t$ of the time interval.
    In both cases, we derive exact
    expressions for the average perimeter of the convex hull and then compare to
    numerical estimates finding excellent agreement.
    Further, we numerically compute the
    full distribution of the perimeter using Markov chain Monte Carlo techniques,
    in both ensembles, probing the far tails of the distribution, up to a
    precision smaller than $10^{-100}$. This also allows us to characterize the
    rare events that contribute to the tails of these distributions.
\end{abstract}
\maketitle


\section{Introduction}
Random walks belong probably to the most thoroughly studied class of
stochastic processes \cite{Chandra_1943,Feller_1968,Hughes_1968} with a wide variety of applications ranging from
finance \cite{Fama1965Random} to biology \cite{berg1972chemotaxis} or online search \cite{page1999PageRank}.
In its simplest form a random walk consists of the sum of independent and identically distributed random jumps,
which converges, in the long time limit when suitably scaled, to Brownian motion (provided the jump distribution
has a finite variance). However, to be a useful model for applications, several variants of this simple model have been introduced and
studied. For example, one may introduce correlations between the steps to model
animal movement \cite{Kareiva1983analyzing,bovet1988spatial} or polymers \cite{Madras2013}. One can also
consider interactions of the walker with its environment to model organisms
that are driven by concentration gradients \cite{codling2008random,Schaefer1973dynamics}.

Here, we will focus on yet another variant, which has found interesting applications in modeling active
matter and active particles. The term ``active particle'' refers to a class of
self-propelled particles which, in contrast to ``passive'' particles such as
Brownian motion, can generate dissipative directed motion by consuming energy directly from their environment \cite{Marchetti_2013,Bechinger_2016,Ram_2017}. Examples of active matter arise in a wide variety of biological and soft matter systems, including bacterial motion \cite{berg1972chemotaxis,Alt1980,Berg2004, Cates2012},  cellular tissue behavior
\cite{tissue}, formation of fish schools \cite{Vicsek, fish} as well as flock of birds \cite{Toner_2005,Kumar_2014}, amongst others. In this context, one of the most studied model is the run-and-tumble
particle (RTP) \cite{Tailleur2008statistical, CT_2015}, initially introduced under the name of the ``persistent random walk'' \cite{Weiss_2002,ML_2017}.
As illustrated in Fig.~\ref{fig_traj}, an RTP performs a ballistic motion along a certain
direction at a constant speed $v_0\ge 0$ (``run'') during a
certain ``time of flight'' $\tau$ after which it ``tumbles'', i.e., chooses a new direction
uniformly at random. Then it performs a new run along this new direction again with speed $v_0$ and so on.
The tumblings occur instantaneously
at random times with constant rate $\gamma$, i.e., the $\tau$ of different runs are independently distributed via
an exponential distribution $p(\tau) = \gamma e^{-\gamma \tau}$.
Despite its
simplicity, this RTP model exhibits complex interesting features such as clustering at boundaries \cite{Bechinger_2016},
non-Boltzmann distribution in the steady state in the presence of a confining
potential~\cite{Tailleur2008statistical, Dhar_18,Sevilla_19,MBE2019,3states_19}, motility-induced phase
separation \cite{CT_2015}, jamming~\cite{SEB2016}, etc.

An interesting question related to the study of animal movements concerns the home range
of an animal, i.e., the two-dimensional territory it covers while searching for food during a
certain period of time~\cite{Worton1995convex}. This is a particularly useful information for ecologists
to decide and design habitat-conservation planning.  A convenient way to estimate this home range is to
construct the convex hull of the trajectory,  i.e., the smallest convex polygon containing every point
visited by the walker (see Fig. \ref{fig_traj}). The perimeter and/or the area of this convex hull provide
quantitative estimates of this home range.

For Brownian motion, the statistics of the convex hull is
a classical problem in probability theory~\cite{Letac1980Expected} and random convex geometry \cite{Hug2013}. Quite
recently, exploiting a beautiful connection to extreme value statistics, exact results have been obtained for the mean
perimeter and mean area for the convex hull of multiple planar Brownian motions \cite{ch1,ch2}, as well as for
the convex hull of a single randomly accelerated particle (also called ``the integrated Brownian motion'') in two dimensions~\cite{RMS2011}. Extensions of these studies of the convex hull for Brownian motion in higher dimensions have also been discussed, mainly in the
mathematics literature~\cite{Eldan2014Volumetric,kabluchko2016intrinsic}. The convex hull of random walks consisting of a finite number of
discrete jumps drawn from some distribution has also been widely studied, both
in the mathematics literature \cite{Kac54, Spitzer56,Snyder,KVZ17a,KVZ17b} and more recently
in physics \cite{GLM17,Claussen2015Convex,schawe2017highdim}. In particular, using
sophisticated sampling methods, higher moments, like the variance, as well as the full distributions
of the perimeter and area of the convex hull of random walks in the plane \cite{Claussen2015Convex}
and in higher dimensions \cite{schawe2017highdim} were recently obtained numerically.

The convex hull of
several variants of the random walk has been studied, including
L\'{e}vy flights \cite{kampf2012convex}, continuous-time random walks \cite{LGE2013},
branching Brownian motion \cite{DMRZ2013} or
self-avoiding random walks \cite{SHM2018,schawe2019true}. However, to the best of our knowledge
nothing is known about
the convex hull of a planar RTP, which, given the wide range of applications of this model,
is clearly an interesting and important observable. The purpose of the present paper is to provide
a detailed study, both analytical and numerical, of the perimeter of the convex hull of an RTP in the plane.

The rest of the paper is organized as follows. In Section~\ref{sec:model} we briefly introduce the
RTP model and present our main results. We proceed with Section~\ref{sec:exact} where we present the details
of our analytical computations. In Section~\ref{sec:num} we outline the numerical methods we have used to simulate RTP, to calculate convex hulls
 as well as how
to obtain the desired distributions for the perimeter of the convex hull even in the tails
down to extremely small probabilities, namely smaller than $10^{-100}$.
Finally, we present our short conclusions in Section~\ref{sec:conclusion}.

\section{Model: The run-and-tumble particle and its convex hull}
\label{sec:model}
\begin{figure}
    \center
    \includegraphics[width=\linewidth]{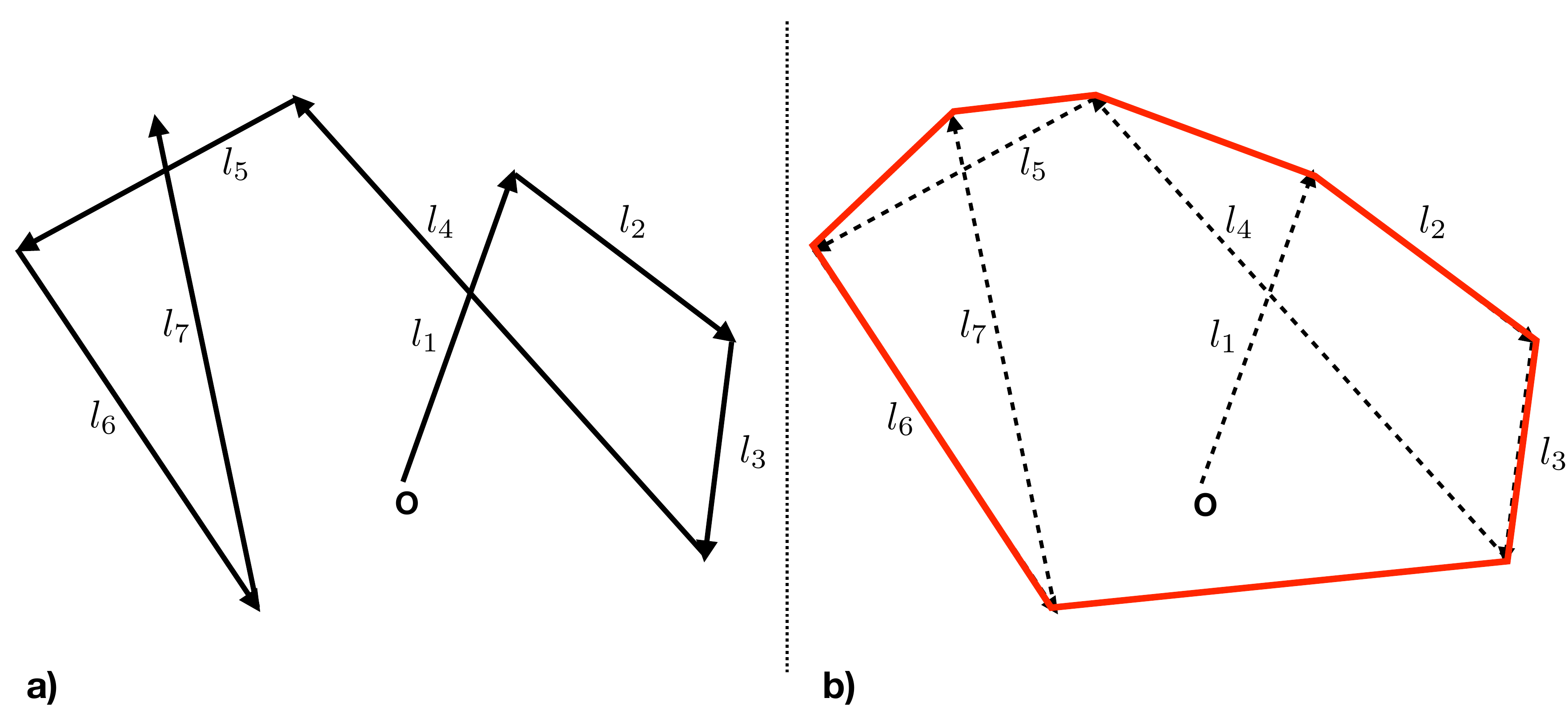}
    \caption{\label{fig_traj}
        {\textbf a)} A typical trajectory of an RTP in two dimensions with $n=7$ steps.
        The particle starts at the origin $O$, chooses a random
        direction and moves ballistically in that direction a distance $l_1=v_0\, \tau_1$ where $v_0$ is
        constant and $\tau_1$ is a random time drawn from the exponential distribution
        $p(\tau)= \gamma\, e^{-\gamma \tau}$. At the end of this first flight, the particle tumbles instantaneously
        and chooses a new random direction and again moves ballistically a distance $l_2= v_0\, \tau_2$
        with $\tau_2$ drawn independently from $p(\tau)= \gamma\, e^{-\gamma\, \tau}$. Then the particle tumbles again
        and so on. The schematic figure shows a trajectory of an RTP with $n$ tumblings, with the starting point
        is counted as a tumble.  {\textbf b)} Same trajectory of an RTP where we have depicted, in red, the convex hull.
    }
\end{figure}

We consider a single run-and-tumble particle in two dimensions (see Fig. \ref{fig_traj}). The dynamics occurs in
continuous time and is defined as follows.
The particle starts at the origin with an initial velocity of fixed magnitude $v_0$ and
chooses a direction at random, i.e., the angle $\phi$ specifying the direction is chosen uniformly
from $[0,2\pi]$.
Subsequently, in a small time $dt$, with probability $\gamma\, dt$, the particle
changes its direction of flight by choosing a new
angle uniformly in $[0,2\pi]$---this is the ``tumbling''. Otherwise
with the complementary probability $1- \gamma\, dt$, the particle continues to move
ballistically with speed $v_0$ in its current direction. Thus $v_0$ and $\gamma$ are the only two parameters in this model. The parameter $\gamma$
denotes the rate of tumbling. The distance covered between two successive tumblings is called a ``run''. We count the starting point $0$ as a tumble and hence the number of tumblings
is the same as the number of runs $n$ and by definition $n\ge 1$.
A typical trajectory for an RTP in the plane is shown in Fig.~\ref{fig_traj}.

We consider two different ensembles: (i) an ensemble where we consider exactly $n$ runs, i.e., we
stop the process exactly after $n$ runs---here the total time spent by the particle fluctuates
from sample to sample while the number of runs $n$ is fixed and (ii) an ensemble where the
total observation time $t$ is fixed---here the number of runs $n$ fluctuates
from sample to sample while the total duration $t$ is fixed. In this paper, we are interested in the convex hull containing the trajectory of the RTP
as an intuitive measure of the geometric size of the RTP. The convex hull is
the smallest convex polygon enclosing a set of points, in this case the set of
points visited by the RTP---imagine a rubber band around the points: it will
contract and become their convex hull (see Fig. \ref{fig_traj}).

For both ensembles, we consider each trajectory of the RTP, draw the unique
convex hull associated to it and compute the mean perimeter of the convex hull
by averaging over all trajectories. Our main analytical results are exact results for the mean
perimeter $\langle L_n \rangle$ in Eq.~(\ref{peri_L_n}) for the ensemble (i)
and $\langle L(t) \rangle$ in Eqs.~(\ref{fixed_t_peri}) and (\ref{Hz_scaling}) for the ensemble (ii).
Besides, we perform simulations of
this model to sample trajectories and measure the distribution of the perimeter
of their hulls. In particular, our numerical results for the mean perimeter are
in very good agreement with our analytical predictions (see Fig. \ref{fig:cmp}).

\section{Exact mean perimeter of the convex hull of a run-and-tumble particle in two dimensions}
\label{sec:exact}

To compute the mean perimeter of the convex hull of a $2d$ RTP, in either
fixed-$n$ or fixed-$t$ ensemble, we use the strategy
developed in Refs.~\cite{ch1,ch2}. It was shown in Refs.~\cite{ch1,ch2} how Cauchy's formula~\cite{Cauchy} for the perimeter
of an arbitrary convex curve in two dimensions can be applied to calculate the mean perimeter of
the convex hull of a generic $2d$ stochastic process. Using this procedure,
the
problem of computing the mean perimeter of the convex hull of an arbitrary $2d$ stochastic process can be mapped to
computing the maximum of the one dimensional component process~\cite{ch1,ch2}. Let us briefly outline the key idea.
Consider an arbitrary convex domain ${\cal D}$ in two dimensions with
its boundary ${\cal C}$ parametrized as $\{{\cal X}(s),{\cal Y}(s)\}$ with $s$ denoting the arc
distance along the boundary contour ${\cal C}$. Cauchy's formula
states that the perimeter of the convex domain ${\cal D}$ is given by
\begin{equation}
    L= \int_0^{2\pi} M(\theta)\, d\theta\, ,
    \label{cauchy.1}
\end{equation}
where $M(\theta)$ is the so called support function
\begin{equation}
    M(\theta)= \max\limits_{s}\left[{\cal X}(s)\cos(\theta)+{\cal Y}(s)\sin(\theta)\right]\, .
    \label{support.1}
\end{equation}
The quantity $M(\theta)$ can be interpreted as follows: Consider the projections of all points of the
boundary curve ${\cal C}$ along the
direction $\theta$ and take the maximum of those projections.

Consider now an arbitrary set of $n$
vertices $\{(X_i, Y_i),\, i=1,2,\ldots, n\}$
in $2d$ (they may represent the positions of a stochastic process in $2d$ at successive times in a given
realization) and construct the convex hull ${\cal C}$ of these vertices. The perimeter of the
convex hull is given by Cauchy's formula in Eq.~(\ref{cauchy.1}). To apply this formula, we need to
first evaluate $\{{\cal X}(s),{\cal Y}(s)\}$ of the convex hull ${\cal C}$ and then compute its maximum over $s$
which seems to be a formidable task.
The key observation of Refs.~\cite{ch1,ch2} that bypasses this step is that
the support function $M(\theta)$ of the convex hull can be obtained directly from the
underlying vertices (without the need to first compute $\{{\cal X}(s),{\cal Y}(s)\}$ of ${\cal C}$ and then
maximizing over $s$) as
\begin{equation}
    M(\theta)=  \max\limits_{1\le i\le n}\left[X_i\, \cos(\theta)+ Y_i\, \sin(\theta)\right]\,.
    \label{support.2}
\end{equation}
Next we average Eq.~(\ref{cauchy.1}) over all realizations of the stochastic process, i.e., over different realizations of
the vertices $\{(X_i,Y_i)\}$ to get
\begin{equation}
    \langle L_n\rangle = \int_0^{2\pi} \langle M(\theta)\rangle \, d\theta\, .
    \label{support.3}
\end{equation}
Moreover, if the $2d$ process is isotropic (e.g.~the RTP process in $2d$ is isotropic),
$\langle M(\theta)\rangle$ can not depend on $\theta$---hence, we may as
well put $\theta=0$. This then simplifies Cauchy's formula and amounts to computing just the
expected maximum of the one-dimensional component
process~\cite{ch1,ch2}
\begin{equation}
    \langle L_n\rangle = 2\, \pi\, \langle M_n\rangle\,  \quad {\rm where}\quad M_n= \max\,\left[X_1,X_2,\ldots,X_n\right]\, .
    \label{max.1}
\end{equation}
This mapping holds for any arbitrary $2d$ isotropic stochastic process. In recent years, this procedure has been
successfully used to compute the mean perimeter for several $2d$ stochastic processes. These include
a single/multiple planar Brownian motions~\cite{ch1,ch2}, planar random acceleration process~\cite{RMS2011},
$2d$ branching Brownian motion with absorption in the context of
epidemic outbreak~\cite{DMRZ2013}, anomalous diffusion processes in $2d$~\cite{LGE2013}, a $2d$ Brownian motion
confined in the half-plane~\cite{CBM12015,CBM22015}, discrete-time $2d$ random walks, L\'evy flights~\cite{GLM17}, etc.
Below we demonstrate that this procedure is also suitable to
compute exactly the mean perimeter of the convex hull of a $2d$ RTP, both in the fixed-$n$
and the fixed-$t$ ensemble.

\subsection{Fixed-$n$ ensemble}

In this ensemble, the total number of runs $n$ of the RTP is fixed, but the duration $t$ fluctuates from sample
to sample. Since the RTP process is isotropic, we can use the general result
in Eq.~(\ref{max.1}). For this, we need to first evaluate the
probability distribution of the coordinates $\{X_1,X_2,\ldots, X_n\}$ of the $x$-component of the
$2d$ RTP. To proceed, consider a particular run, say the $i$-th run in Fig.~\ref{fig_traj}.
The length of the run is $l_i= v_0 \tau_i$, where $\tau_i$ is distributed exponentially
$p(\tau_i)= \gamma\, e^{-\gamma\, \tau_i}$. When projected along the $x$-axis, this corresponds
to an increment $x_i= l_i\, \cos(\phi_i)$ in the $x$-direction, where the angle $\phi_i$ is
distributed uniformly over $\phi_i\in[0,2\pi]$. Let us write, $x_i= v\, \tau_i$, where
$v= v_0\, \cos(\phi_i)$. Given the uniform distribution of $\phi_i$, it is easy to
compute the distribution $P(v)$ of $v$ using $P(v)dv= d\phi_i/(2\pi)$ and we get
\begin{equation}
    P(v)= \frac{1}{\pi\, \sqrt{v_0^2-v^2}}\, ; \quad \,\, -v_0\le v\le v_0\, .
    \label{pv1_n}
\end{equation}
Consequently, the joint distribution $p(x,\tau)$ of the increment $x_i=v\,\tau_i$ (along the $x$-direction)
and the duration $\tau_i$ of the $i$-th run
is given by
\begin{equation}
    \fl 
    p(x,\tau)= {\rm Prob.}\left[x_i=x,\, \tau_i=\tau\right]= \int_{-v_0}^{v_0} dv\, \delta(x- v\tau)\,
    \frac{1}{\pi\, \sqrt{v_0^2-v^2}}\, \gamma\, e^{-\gamma\, \tau}\, .
    \label{pxt1_n}
\end{equation}
Integrating over $v$ gives
\begin{equation}
    p(x,\tau)= \frac{\gamma\, e^{-\gamma\, \tau}}{\pi\, \sqrt{v_0^2\tau^2-x^2}}\, \theta\left(\tau- \frac{x}{v_0}\right)\, ,
    \label{pxt2_n}
\end{equation}
where $\theta(z)$ is the standard Heaviside step function: $\theta(z)=1$ if $z>0$ and $\theta(z)=0$ if $z<0$.

Integrating further over $\tau$, one sees that the increment $x_i$ is distributed via the marginal probability density
\begin{equation}
    \fl 
    f(x)= {\rm Prob.}[x_i=x]= \int_0^{\infty} p(x,\tau)\, d\tau=
    \frac{\gamma}{\pi}\, \int_{x/v_0}^{\infty}
    \frac{ d\tau\, e^{-\gamma\, \tau}}{\sqrt{v_0^2\tau^2-x^2}}=
    \frac{\gamma}{\pi\, v_0}\, K_0\left(\frac{\gamma |x|}{v_0}\right)\, ,
    \label{fx2_n}
\end{equation}
where $K_\nu(z)$ is the modified Bessel function with index $\nu$. One can check easily that $f(x)$ is normalized to unity:
$\int_{-\infty}^{\infty} f(x)dx=1$. Furthermore, the variance $\sigma^2$ of this distribution is finite and
is given by
\begin{equation}
    \sigma^2= \int_{-\infty}^{\infty} x^2\, f(x)\, dx= \frac{2}{\pi}\left(\frac{v_0}{\gamma}\right)^2\,
    \int_0^{\infty} y^2\, K_0(y)\, dy= \left(\frac{v_0}{\gamma}\right)^2\, .
    \label{varx_n}
\end{equation}

Thus, each run of the $2d$ RTP gives rise to an independent increment $x$ in the $x$-direction which
is distributed via the symmetric, continuous, normalized to unity probability distribution function (PDF)
$f(x)$ in Eq.~(\ref{fx2_n}). Thus the projected $x$-component process is just a one dimensional
random walk, $X_i= X_{i-1}+ x_i$ with independent and identically distributed (i.i.d.) increments
$x_i$, each drawn from $f(x)$ in Eq.~(\ref{fx2_n}), and starting at $X_0=0$. Hence, we need to now compute the
expected maximum $\langle M_n\rangle$ of this reduced one dimensional random walk process of $n$ steps.
This can be conveniently computed using a formula originally due to Kac~\cite{Kac54} (see also
Ref.~\cite{Spitzer56})
\begin{equation}
    \langle M_n\rangle = \frac{1}{2}\, \sum_{m=1}^n \frac{\langle |X_m|\rangle}{m} \, .
    \label{kac_n}
\end{equation}
To compute $\langle |X_m|\rangle$, we need to know the distribution $P_m(X)= {\rm Prob.}(X_m=X)$
of the position of the $1d$ random walker at step $m$. This can be easily computed as follows.
Clearly, $X_m=x_1+x_2+\ldots +x_m$ where $x_i$ are i.i.d.\ random variables each drawn from $f(x)$.
Hence
\begin{equation}
    P_m(X)= \int_{-\infty}^{\infty} \delta\left(X- \sum_{i=1}^m x_i\right)\, \prod_{i=1}^n f(x_i)\, dx_i \, .
    \label{Pmx_n}
\end{equation}
Taking the Fourier transform gives
\begin{equation}
    {\tilde P}_m(k)= \int_{\infty}^{\infty} P_m(X)\, e^{i\, k\, X}\, dX= \left[{\tilde f}(k)\right]^m\, ,
    \label{Pmk_n}
\end{equation}
where the Fourier transform of the jump distribution $f(x)$ in Eq.~(\ref{fx2_n}) turns out to be rather simple
\begin{equation}
    {\tilde f}(k) =  \frac{\gamma}{\pi\, v_0}\, \int_{-\infty}^{\infty} e^{i\,k\,x}\,
    K_0\left(\frac{\gamma |x|}{v_0}\right)\, dx=  \frac{1}{\sqrt{1+ \frac{v_0^2\, k^2}{{\gamma}^2}}}\, .
    \label{fk_n}
\end{equation}
Substituting this in Eq.~(\ref{Pmk_n}) and inverting the Fourier transform (fortunately it can be done explicitly)
we get
\begin{eqnarray}
    P_m(X) & = &\int_{-\infty}^{\infty}
    \left[1+ \frac{v_0^2\, k^2}{\gamma^2}\right]^{-m/2}\, e^{-i\,k\,X}\, dk \nonumber \\
    &=& \frac{\gamma\, 2^{(1-m)/2}}{v_0\, \sqrt{\pi}\, \Gamma(m/2)}\,
    \left(\frac{\gamma\, |X|}{v_0}\right)^{(m-1)/2}\, K_{(1-m)/2}\left(\frac{\gamma\, |X|}{v_0}\right)\, .
    \label{PmX1_n}
\end{eqnarray}
where $\Gamma(z)$ is the standard Gamma function.

From this exact PDF of $X_m$ in Eq.~(\ref{PmX1_n}), one can easily compute $\langle |X_m|\rangle$
\begin{eqnarray}
    \fl 
    \langle |X_m|\rangle= \int_{-\infty}^{\infty} |X|\, P_m(X)\, dX&=& \frac{v_0}{\gamma}\,
    \frac{2^{(3-m)/2}}{\sqrt{\pi}\, \Gamma(m/2)}\, \int_0^{\infty} y^{(m+1)/2}\, K_{(1-m)/2}(y)\, dy \\
    &=& \frac{2\,v_0}{\gamma}\, \frac{\Gamma\left(\frac{1+m}{2}\right)}{\sqrt{\pi}\, \Gamma(m/2)}
    \label{X_m1_n} \;.
\end{eqnarray}
Substituting this result in Kac's formula (\ref{kac_n}) and using Eq.~(\ref{max.1}) gives,
for any $n\ge 1$,
\begin{equation}
    \langle L_n \rangle = \frac{v_0\sqrt{\pi}}{\gamma}\, \sum_{m=1}^n
    \frac{\Gamma\left(\frac{m}{2}+\frac{1}{2}\right)}{\Gamma\left(\frac{m}{2}+1\right)}\, .
    \label{fixed_n_peri}
\end{equation}
This sum can be done explicitly and we get
\begin{equation}
    \fl 
    \langle L_n \rangle = \frac{v_0}{\gamma}\, \left[-(\pi+2) + 2\,\sqrt{\pi}\,
    \left\{ \frac{\Gamma\left(2+\lfloor{\frac{n-1}{2}}\rfloor\right)}{
    \Gamma\left(\frac{3}{2}+\lfloor{\frac{n-1}{2}}\rfloor\right)}
    + \frac{\Gamma\left(\frac{3}{2}+\lfloor{\frac{n}{2}}\rfloor\right)}{
    \Gamma\left(1+\lfloor{\frac{n-1}{2}}\rfloor\right)}
    \right\} \right] \, ,
    \label{peri_L_n}
\end{equation}
where $\lfloor{z}\rfloor$ denotes the integer part of $z$.
It is easy to check that for large $n$ one gets the asymptotic behavior
\begin{equation}
    \langle L_n \rangle \to \frac{v_0}{\gamma}\, \sqrt{8\pi n}\, \quad {\rm as}\,\, n\to \infty \, .
    \label{fixed_n_peri_asymp}
\end{equation}
Using $\sigma= v_0/\gamma$ from Eq.~(\ref{varx_n}), we see that for large $n$, $\langle L_n\rangle \to
\sigma \, \sqrt{8\,\pi\, n}$ which coincides with the mean perimeter of the convex hull of a discrete
two dimensional random walk of $n$ steps for any jump distribution with a finite variance $\sigma^2$~\cite{GLM17}.

\subsection{Fixed-$t$ ensemble}

Here, we consider the total duration $t$ fixed, but the number of runs $n$ during $t$ may fluctuate from
sample to sample and hence $n$ is a random variable.
Note that the duration $\tau_n$ of the last interval (i.e., following the
$n$-th tumbling) traveled by the particle
before the epoch $t$ is yet to be complete. Consequently, its distribution
is $e^{-\gamma\, \tau_n}$ (denoting the probability of no tumbling during the interval of duration $\tau_n$)
is not normalized to unity. This is in contrast to the already completed preceding intervals
each of which is distributed independently
according to the normalized distribution $p(\tau)= \gamma\, e^{-\gamma\, \tau}$. For each of these
preceding intervals, the joint distribution of the $x$-component and the duration of the
interval is given by $p(x,\tau)$ in Eq.~(\ref{pxt2_n}) and each interval is independent of the other.
The corresponding joint distribution for the last (incomplete run) interval, in contrast, is different by a factor
$1/\gamma$
\begin{equation}
    p_{\rm last}(x,\tau)=  \frac{e^{-\gamma\, \tau}}{\pi\, \sqrt{v_0^2\tau^2-x^2}}\, \theta\left(\tau- \frac{x}{v_0}\right)
    = \frac{1}{\gamma}\, p(x,\tau)\, ,
    \label{joint_last1_t}
\end{equation}
where $p(x,\tau)$ is given in Eq.~(\ref{pxt2_n}). Note however that the sum of the durations of the $(n-1)$ completed
runs and the last incomplete run is fixed to be $t$. Hence, the grand joint distribution of the $x$-increments $\{x_i\}$,
their associated durations $\{\tau_i\}$ (where $i=1,2,\ldots, n$) and the number of runs $n$ is given by
\begin{eqnarray}
    \fl 
    P\left[\{x_i\},\, \{\tau_i\}, \, n|t\right]&=& \frac{1}{\gamma}\,\left[ \prod_{i=1}^n p(x_i,\tau_i)\right]\,
    \delta\left(\sum_{i=1}^n \tau_i-t\right) \nonumber \\
    &=& \frac{1}{\gamma}\,\left[ \prod_{i=1}^n
    \frac{\gamma\, e^{-\gamma\, \tau_i}}{\pi\, \sqrt{v_0^2\tau^2-x_i^2}}\, \theta\left(\tau_i- \frac{x_i}{v_0}\right)\,
    \right]\,
    \delta\left(\sum_{i=1}^n \tau_i-t\right)\, .
    \label{grand_joint1_t}
\end{eqnarray}
The presence of the delta function on the right hand side (rhs) of Eq.~(\ref{grand_joint1_t}) naturally signals
that it is convenient to work in the Laplace space conjugate to $t$. Taking the
Laplace transform of Eq.~(\ref{grand_joint1_t})
with respect to $t$ and integrating over the $\tau_i$ variables, we obtain the
Laplace transform of the marginal joint distribution $P\left[\{x_i\},n|t\right]$ in a factorized form
\begin{eqnarray}
    \fl 
    \int_0^{\infty} P\left[\{x_i\},\, n|t\right]\, e^{-s\, t}\, dt &=&
    \frac{1}{\gamma}\,\left[ \prod_{i=1}^n
    \int_{x_i/v_0}^{\infty}
    \frac{\gamma\, e^{-(\gamma+s)\, \tau_i}}{\pi\, \sqrt{v_0^2\tau^2-x_i^2}}\, d\tau_i \right] \nonumber \\
     &=& \frac{1}{\gamma}\, \left[ \prod_{i=1}^n \frac{\gamma}{v_0\,\pi}\, K_0\left(\frac{(\gamma+s)\,|x_i|}{v_0}\right)\right]\, ,
    \label{joint_xn1_t}
\end{eqnarray}
where we used the result from Eq.~(\ref{fx2_n}). To proceed further, it is convenient to define a normalized (to unity)
PDF $f_s(x)$ parametrized by $s$ as follows
\begin{equation}
    f_s(x)= \frac{ K_0\left(\frac{(\gamma+s)\,|x|}{v_0}\right)}{\int_{-\infty}^{\infty}
    K_0\left(\frac{(\gamma+s)\,|x|}{v_0}\right)\, dx}= \frac{\gamma+s}{\pi\, v_0}\, K_0\left(\frac{(\gamma+s)\,
    |x|}{v_0}\right)\, .
\label{fsx1_t}
\end{equation}
In terms of this normalized PDF $f_s(x)$, we can rewrite Eq.~(\ref{joint_xn1_t}) in a convenient form as
\begin{equation}
    \int_0^{\infty} P\left[\{x_i\},\, n|t\right]\, e^{-s\, t}\, dt=
    \frac{1}{\gamma}\, \left(\frac{\gamma}{\gamma+s}\right)^n\, \prod_{i=1}^n f_s(x_i)\, .
    \label{joint_xn2_t}
\end{equation}
We next invert the Laplace transform formally as
\begin{equation}
    P\left[\{x_i\},\, n|t\right]= \int_{\Gamma} \frac{ds}{2\pi i}\, e^{s\,t}\,
    \frac{1}{\gamma}\, \left(\frac{\gamma}{\gamma+s}\right)^n\, \prod_{i=1}^n f_s(x_i)\, ,
    \label{joint_xn3_t}
\end{equation}
where $\Gamma$ denotes a vertical Bromwich contour (to the right of all singularities of the integrand)
in the complex $s$ plane. Note that the increments $x_i$ are correlated since the rhs of
Eq.~(\ref{joint_xn3_t}) does not factorize.

So, once again, the projected $x$-increments $\{x_i\}$ ($i=1,2\ldots, n$) form a random walk process
in $1d$ where the position $X_i$ of the walker evolves as $X_i=X_{i-1}+x_i$, starting from $X_0=0$.
However, unlike in the fixed-$n$ ensemble, the increments in the fixed-$t$ ensemble are not independent random
variables, but are correlated as in Eq.~(\ref{joint_xn3_t}). To compute the mean perimeter $\langle L(t)\rangle$
of the convex hull for fixed $t$ using Eq.~(\ref{max.1}), we need to compute the expected maximum of the
$1d$ process $X_i$ with fixed $t$ whose increments
and number of steps $n$ are jointly distributed via Eq.~(\ref{joint_xn3_t}), and finally sum over all $n$.
It is convenient to define the following quantity
\begin{eqnarray}
    &&\fl 
    Q(M,n|t)= \int_{-\infty}^{\infty} dx_1\ldots \int_{-\infty}^{\infty}\, dx_n\,
    {\rm Prob.}\left[X_1<M, X_2 < M,\, \ldots, X_n<M, n|t\right] \;, \label{cumul_Mn_t} \\
   &&{\rm with} \qquad X_i = \sum_{j=1}^i x_j  \;,\label{def_Si}
\end{eqnarray}
that denotes the joint probability that the maximum of the random walk process is less than $M$ and the
number of steps is exactly $n$, for a given $t$. Taking a derivative with respect to $M$ gives
the joint PDF of $M$ and $n$: $P(M,n|t)= \partial_M Q(M,n|t)$. Using Eq.~(\ref{max.1}) we then have
\begin{equation}
    \langle L(t)\rangle= 2\,\pi\, \sum_{n=1}^{\infty} \int_{0}^{\infty} dM\, M\, \partial_M Q(M,n|t)\, .
    \label{lt_max_t}
\end{equation}
Using the joint distribution in Eq.~(\ref{joint_xn3_t}) and the definition $Q(M,n|t)$ in Eq.~(\ref{cumul_Mn_t})
we then get formally
\begin{equation}
    \langle L(t)\rangle= 2\,\pi\, \sum_{n=1}^{\infty}\, \int_{\Gamma} \frac{ds}{2\pi i}\, e^{s\,t}\,
    \frac{1}{\gamma}\, \left(\frac{\gamma}{\gamma+s}\right)^n\, \langle M_s(n)\rangle\, ,
    \label{lt_max1_t}
\end{equation}
with
\begin{equation}
    \langle M_s(n)\rangle= \int_0^{\infty} dM\, M\, \partial_M Q_s(M,n)\,
    \label{Msn_t}
\end{equation}
where
\begin{eqnarray}
    &&\fl 
    Q_s(M,n)= \int_{-\infty}^{\infty}\ldots \int_{-\infty}^{\infty} \left[\prod_{i=1}^n f_s(x_i)\, dx_i\right]
    {\rm Prob.}\left[Y_1<M, \, Y_2<M,\,  \ldots, Y_n<M\right] \\
    &&{\rm with} \qquad Y_i = \sum_{j=1}^i x_j  \; .\label{def_Yi}
\label{Qsmn1_t}
\end{eqnarray}

Thus, $Q_s(M,n)$ in Eq.~(\ref{Qsmn1_t}) can be interpreted as the cumulative distribution of the
maximum of an auxiliary $1d$ random walk process $\{Y_i\}$ of $n$-steps,
$Y_{i}=Y_{i-1}+x_i$, with i.i.d.\ increments $x_i$ each distributed via $f_s(x)$
given in Eq.~(\ref{fsx1_t}). Consequently, $\langle  M_s(n)\rangle$
in Eq.~(\ref{Msn_t}) is simply the expected maximum of this auxiliary $n$-step $1d$ process $\{Y_i\}$
parametrized by $s$.
However, as mentioned before, the expected maximum of any $1d$ random walk process with i.i.d.\ increments
can be computed using Kac's formula in Eq.~(\ref{kac_n}), provided we can compute the expected absolute value
of the process $\langle |Y_m|\rangle $ at step $m$. Then, we get
\begin{equation}
    \langle M_s(n)\rangle= \frac{1}{2}\, \sum_{m=1}^n \frac{\langle |Y_m|\rangle}{m}
    \label{kac_t}
\end{equation}
The average of $|Y_m|$ can be computed explicitly as in the
case of the fixed-$n$ ensemble discussed before.
Using $Y_m= x_1+x_2+\ldots x_m$ where $x_i$ are i.i.d.\ variables
each drawn from the PDF $f_s(x)$, the PDF $P_m(Y)$ of $Y_m$ can be computed in terms of the
Fourier transform of the jump distribution $f_s(x)$ as in Eq.~(\ref{Pmk_n})
\begin{equation}
    P_m(Y)= \int_{-\infty}^{\infty}
    \left[{\tilde f}_s(k)\right]^m \, e^{-i\,k\,Y}\, \frac{dk}{2\pi}\, ,
    \label{PmY1_t}
\end{equation}
where
\begin{equation}
    {\tilde f}_s(k)= \int_{-\infty}^{\infty} f_s(x)\, e^{i\,k\,x}\, dx= \frac{1}{\sqrt{\left(
    \frac{v_0}{\gamma+s}\right)^2\, k^2+1}}\, .
    \label{fsk_t}
\end{equation}
We have used the explicit expression of $f_s(x)$ from Eq.~(\ref{fsx1_t}) and the
result in Eq.~(\ref{fk_n}) where we replaced $\gamma$ by $\gamma+s$.
Substituting Eq.~(\ref{fsk_t}) in Eq.~(\ref{PmY1_t}) and performing the
integral (as in Eq.~(\ref{PmX1_n}) with $\gamma$ replaced $\gamma+s$) we get
\begin{equation}
    \fl 
    P_m(Y)= \frac{(\gamma+s)\, 2^{(1-m)/2}}{v_0\, \sqrt{\pi}\, \Gamma(m/2)}\,
    \left(\frac{(\gamma+s)\, |Y|}{v_0}\right)^{(m-1)/2}\, K_{(1-m)/2}\left(\frac{(\gamma+s)\, |Y|}{v_0}\right)\, .
    \label{PmY2_t}
\end{equation}
This then gives
\begin{equation}
    \langle |Y_m|\rangle= \int_{-\infty}^{\infty} |Y|\, P_m(Y)\, dY=
    \frac{2\,v_0}{\gamma+s}\, \frac{\Gamma\left(\frac{1+m}{2}\right)}{\sqrt{\pi}\, \Gamma(m/2)}\, .
    \label{Y_m1_t}
\end{equation}
Substituting this result in Eq.~(\ref{kac_t}) we get
\begin{equation}
    \langle M_s(n)\rangle= \frac{v_0}{(\gamma+s)\,\sqrt{4\pi}}\,
    \sum_{m=1}^n \frac{\Gamma\left(\frac{m}{2}+\frac{1}{2}\right)}{\Gamma\left(\frac{m}{2}+1\right)}\,
    \label{Msn2_t}
\end{equation}
and consequently Eq.~(\ref{lt_max1_t}) yields
\begin{equation}
    \fl 
    \langle L(t)\rangle= v_0\, \sqrt{\pi}\,\sum_{n=1}^{\infty} \gamma^{n-1}\,
    \left[\sum_{m=1}^n \frac{\Gamma\left(\frac{m}{2}+\frac{1}{2}\right)}{\Gamma\left(\frac{m}{2}+1\right)}\right]\,
    \int_{\Gamma} \frac{ds}{2\pi i}\, e^{s\,t}\, \frac{1}{(\gamma+s)^{n+1}}\, .
    \label{lt3_t}
\end{equation}
The Bromwich integral can now be trivially performed using
\begin{equation}
    \int_{\Gamma} \frac{ds}{2\pi i}\, e^{s\,t}\, \frac{1}{(\gamma+s)^{n+1}}= \frac{t^n\, e^{-\gamma\, t}}{\Gamma(n+1)}\, .
    \label{lt4_t}
\end{equation}
This then gives us the formula for $\langle L(t)\rangle$ in the fixed-$t$ ensemble
\begin{equation}
    \langle L(t)\rangle= \frac{v_0\,\sqrt{\pi}}{\gamma}\, e^{-\gamma\,t} \sum_{n=1}^{\infty} \frac{(\gamma\, t)^n}{n!}\,
    \left[\sum_{m=1}^n \frac{\Gamma\left(\frac{m}{2}+\frac{1}{2}\right)}{\Gamma\left(\frac{m}{2}+1\right)}\right]\, .
    \label{lt5_t}
\end{equation}

The result in Eq.~(\ref{lt5_t}) is valid at any time $t$. Furthermore, it turns out that the
double summation can be performed explicitly and $\langle L(t)\rangle$ can be written
in a scaling form
\begin{equation}
    \langle L(t)\rangle = \frac{v_0}{\gamma}\, H(\gamma\, t)\, ,
    \label{fixed_t_peri}
\end{equation}
where the scaling function $H(z)$ is given exactly by
\begin{eqnarray}
    \fl 
    H(z)= e^{-z}\,\left[2- (\pi+2)\,e^z + 2\,z + \pi\,(1+z)\, (I_0(z)+L_0(z)) + \pi\, z\, (I_1(z)+L_1(z)) \right]
    \label{Hz_scaling}
\end{eqnarray}
In Eq.~(\ref{Hz_scaling}), $I_\nu(z)$ is the standard modified Bessel functions of index $\nu$,
while $L_\nu(z)$ denotes the Struve function of index $\nu$ defined as
\begin{equation}
    L_\nu(z)= \left(\frac{z}{2}\right)^{\nu+1}\, \sum_{k=0}^{\infty}
    \frac{1}{\Gamma\left(\frac{3}{2}+k\right)\Gamma\left(\frac{3}{2}+ k+ \nu\right)}\,
    \left(\frac{z}{2}\right)^{2k}\, .
    \label{struve.1}
\end{equation}

\begin{figure}
    \centering
    \includegraphics[width=0.8\textwidth]{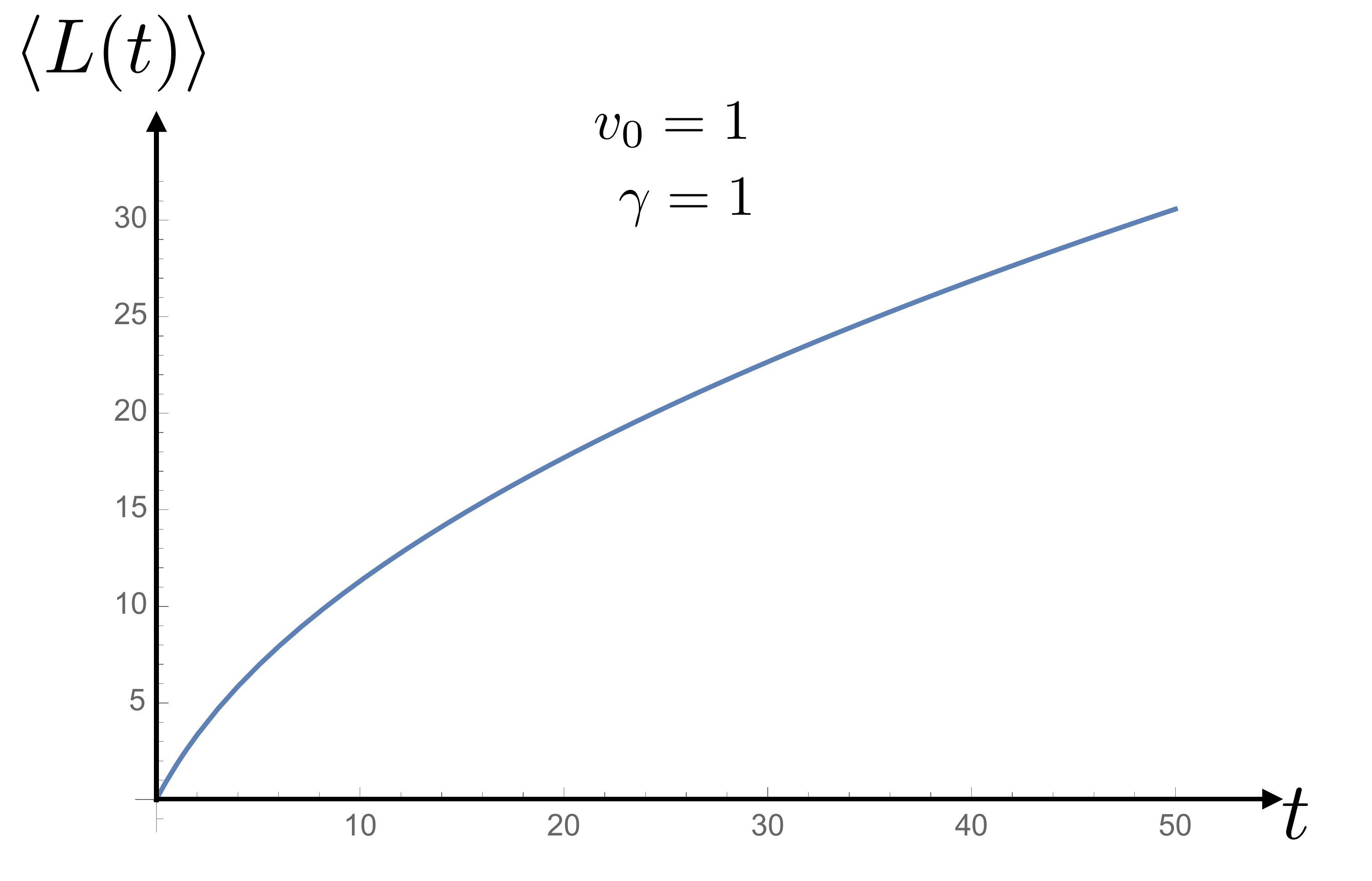}
    \caption{\label{fig_Plot_Lt}
        The mean perimeter of the convex hull of an RTP $\langle L(t)\rangle$ in Eqs. (\ref{fixed_t_peri}) and
        (\ref{Hz_scaling}), plotted as a function of $t$ (using Mathematica) for parameter values $v_0=1$ and $\gamma=1$.
    }
\end{figure}

The function $H(z)$ in Eq.~(\ref{Hz_scaling}) can be plotted using
\emph{Mathematica}---a plot of $\langle L(t)\rangle$ vs.~$t$ for
$v_0=1$ and $\gamma=1$ is shown in Fig.~\ref{fig_Plot_Lt}.
The scaling function $H(z)$ has the
following asymptotic behaviors
\begin{equation}\label{Hz_asymp}
    H(z)= \cases{ 2\,z + \frac{\pi-4}{4}\, z^2 + O(z^3)\, , & $\mathrm{as }\,\, z\to 0$  \;,\cr
    \sqrt{8\,\pi\, z} - (\pi+2) + O\left(\frac{1}{\sqrt{z}}\right)\, , & $\mathrm{as } \,\, z\to \infty$ \;.
    }
\end{equation}
Consequently, the mean perimeter $\langle L(t)\rangle$ initially grows ballistically
for $t\ll \gamma^{-1}$ and eventually for $t\gg \gamma^{-1}$, it crosses over to the diffusive behavior
\begin{eqnarray}\label{Lt_asymp}
    \langle L(t) \rangle \to \cases{ 2\, v_0\, t + O(t^2)\, ,&${\rm for}\,\, t\ll \gamma^{-1}$  \cr
    \frac{v_0}{\gamma}\left[\sqrt{8\pi \gamma t} - (\pi+2) + O\left(\frac{1}{\sqrt{t}}\right)\right] &${\rm for}
    \, \, t\gg \gamma^{-1}$\, .
    }
\end{eqnarray}
We remark that in the limit $v_0\to \infty$, $\gamma\to \infty$, but with the ratio $v_0^2/(2\gamma)=D_{\rm eff}$ kept fixed,
it is well known that the RTP converges to a standard Brownian motion with an effective diffusion constant $D_{\rm eff}$.
In this limit, we get from the second line of Eq.~(\ref{Lt_asymp}), $\langle L(t)\rangle \to \sqrt{16\,\pi\,D_{\rm eff}\,t}$
and thus coincides with the well known result for the mean perimeter of the convex hull of a
$2d$ Brownian motion of duration $t$ and diffusion constant $D_{\rm eff}$ (see e.g.~the review~\cite{ch2}).
Thus, at short times $t\ll \gamma^{-1}$, the behavior of $\langle L(t)\rangle$ for the RTP is clearly
different from that of a passive (ordinary) Brownian motion. However, at late times, the signature of activity is present
not in the leading term (which effectively behaves as in the passive Brownian motion), but
in the subleading (second term in the last line of Eq.~(\ref{Lt_asymp})) nontrivial constant
$-(\pi+2)v_0/\gamma$.

To summarize, our principal results in this section concern the exact formulae for the mean perimeter of the
convex hull of an RTP, both in the ensemble of fixed number $n$ of runs (as given in Eq.~(\ref{peri_L_n}) which is valid for all $n$
) and in the fixed time $t$ ensemble (given in Eq.~(\ref{fixed_t_peri}) which is
valid for all $t$). Later, in subsection (\ref{Num_Verific}) we compare the high precision simulation results to these
exact formulae for the mean perimeter derived in this section. In Figs.~\ref{fig:cmp:n} and \ref{fig:cmp:t}
we compare the simulation results to the analytical formula for these two ensembles and find excellent agreement.

\section{Numerically estimated mean perimeter and higher moments}
\label{sec:num}
Here we present the numerical part of this study. This section is split in three
parts: Subsection \ref{sec:num_methods} is a rather technical part of interest for
the reader wanting to reproduce the results, which may be safely skipped when
only interested in the results. Subsection \ref{Num_Verific} verifies the
analytic results and subsection \ref{sec:num_dist} studies the full distribution
of the perimeter of the RTP's convex hull.

\subsection{Methods}
\label{sec:num_methods}
Modeling the RTP in simulation is straightforward. One draws random
directions $\phi_i$ uniformly from $[0,2\pi]$
and the duration of the runs, i.e., the run times between tumble events,
$\tau_i$ from the exponential distribution
\begin{equation}
    \label{eq:exponential_dist}
    p(\tau) = \gamma \rme^{-\gamma \tau},
\end{equation}
where the latter task is easily achieved using the standard
\emph{inversion method}  \cite{practical_guide2015}.
This is repeated, depending on the ensemble we want to simulate, until we
either have drawn $n$ random numbers of each or
until the total time $\sum_i \tau_i \ge t$, in which case the last waiting time
is truncated to result in an equality.

The convex hull of point sets in a $2d$ plane is easy to obtain using
well established methods, such as Andrew's Monotone Chain
algorithm \cite{Andrew1979Another}, which exploit the fact that
a convex hull in $2d$ is defined by the order of its vertices and find the hull
of a point set of size $n$ in time $O(n\log(n))$. For our purposes a further
speedup can be achieved by preprocessing the point set with Akl's heuristic \cite{Akl1978Fast}.
The main idea behind this heuristic is that all points which are inside the convex hull of a subset, are
also inside of the convex hull of the actual set and therefore not part of the
convex hull. Using the points with minimal and maximal $x$- and $y$-coordinates,
as well as points with extreme $x+y$ and $x-y$ as vertices for the subset hull,
usually allows to discard the majority of points in time $O(n)$ such that the exact
algorithm can operate on a much smaller point set.

Given the set of $n$ points in $2d$, let $m\le n$ denote the number of vertices
of the associated convex hull.
Given these $m$ vertices in order by their Cartesian
coordinates $\{(X_i, Y_i)\}$, the calculation of its perimeter is trivial
\begin{equation}
    L = \sum_{i=0}^{m-1} \sqrt{(X_i-X_{i+1})^2 + (Y_i-Y_{i+1})^2},
\end{equation}
with cyclical indexes, i.e., $X_0 = X_m$.

We employ two types of simulation to obtain our numerical results. To estimate
the mean values and variances, we use \emph{simple sampling}.
That is, we generate
$10^5$ independent realizations of the RTP ensemble naively, construct their
convex hulls and calculate the perimeter of each. These samples can be used to
estimate the mean and variance.
To obtain the distributions of the perimeter close to its typical values, it is sufficient to create a
histogram from the collected samples. But note that the
tails of this histogram will only contain events which occur with a probability,
for our sample size, of around $10^{-5}$ or higher.

To reach the tails of the distribution containing extremely rare events with
probabilities of, say, $10^{-100}$, we need to employ a more sophisticated
sampling method. Therefore, we use a \emph{Monte Carlo} (MC)
method based on Markov chains.
Each state of the Markov chain consists
of one realization of an RTP trajectory.
The general idea is to use a Markov chain
to generate realizations whose appearance probability is weighted with a
known weight depending on the observable of interest, here the perimeter $L$.
This weight has to exhibit a free parameter, which is used to bias the
realizations
towards different regions in $L$-space, such that good statistics can
be sampled for atypical values of $L$. Afterwards the knowledge about the bias
is used to calculate the wanted distribution $P(L)$, from the biased
realizations.

\begin{figure}[htb]
    \centering

    \includegraphics[scale=1]{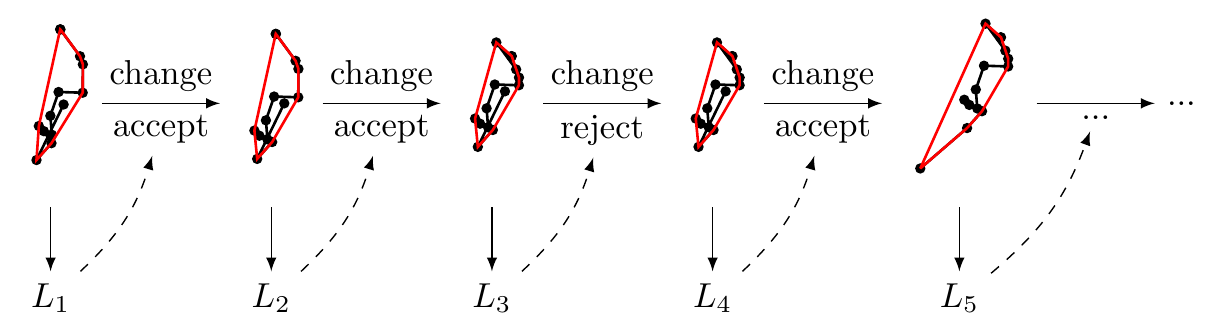}

    \caption{\label{fig:mcmc}
        Sketch of the Markov chain of RTP realizations. At each step a change
        is proposed and either accepted, leading to a new entry in the chain
        or rejected, leading to a duplicate of the old entry in the chain.
        In this example a negative temperature leading to larger than usual
        hulls is used for the fixed-$n$ ensemble with $n=12$.
    }
\end{figure}

An example of a Markov chain of realizations of RTP trajectories is sketched
in Fig.~\ref{fig:mcmc}. To transition
from the current state to the next state of the chain, we have to define
a \emph{change move}.
To decide for a simple yet  efficient  change move,
we have to look at the representation of an RTP trajectory
realization. Here we define it by a tuple of times between tumble events
$(\tau_1, \tau_2, \ldots)$ and a tuple of directions chosen at the tumble events
$(\phi_1, \phi_2, \ldots)$. A change move is constructed by replacing one
random entry of the tuples by a new random waiting time $\tau_i'$ from
the exponential distribution $p(\tau)$ or by a new random direction
from the corresponding uniform distribution $\phi_i'$. Note that if this change
was always accepted, it would lead to a Markov chain whose entries would
be distributed uniformly over all RTP trajectory realizations.

To introduce the bias, we use the classical Metropolis algorithm
\cite{metropolis1953equation} which generates in its original formulation
realizations of the canonical ensemble at a given temperature.
To foster intuition, in the canonical ensemble at low
temperatures one encounters realizations near the ground state, i.e., very low
energies, and at high temperatures realizations with high energies.
Here, we identify the observable of interest $L$ with the energy appearing in
the original Metropolis algorithm and can use the temperature $\Theta$ as a
free parameter to bias the resulting distribution.
Therefore we accept each proposed change with the probability $p_\mathrm{acc} =
\min\left\{ 1, e^{-\Delta L / \Theta} \right\},$ where
$\Delta L$ is the difference of the perimeter in the current and the
proposed state.

If a proposed change is rejected, the current step will be repeated in the
Markov chain. This way a sample of the states appearing in the Markov chain
will eventually be distributed such that realizations $\mathcal{C}$ appear
distributed according to $Q_\Theta(\mathcal{C}) = \frac{1}{Z(\Theta)} Q(\mathcal{C}) e^{-L(\mathcal{C})/\Theta}$,
where $Q(\mathcal{C})$ is the distribution of the realizations, i.e., when
drawing uniformly from all realizations of RTP trajectories.
The quasi-Boltzmann factor
biases the samples to large or small values of $L$ depending on the artificial
temperature $\Theta$, which can assume positive and negative values. Note that
negative $\Theta$ will lead to realizations with atypically large perimeter.

Summing the distribution $Q(\mathcal{C})$ over all realizations with the
same perimeter results in the wanted distribution $P(L)$, and after some
elemental algebra we obtain the relation
\begin{equation}
    P(L) = e^{L/\Theta} Z(\Theta) P_\Theta(L)
\end{equation}
between the distribution we measure in the biased ensembles $P_\Theta(L)$
and the wanted distribution $P(L)$.
The ratios of the two unknown constants $Z(\Theta_i)$ and $Z(\Theta_j)$ can be
obtained by enforcing that the two estimates of the corresponding biased
distributions $P_\Theta(L)$ must coincide in overlapping regions.
The absolute value of $Z(\Theta)$ is obtained by normalization
of the entire distribution. This way,
the wanted distribution $P(L)$ is obtained over a very large range.

Due to the small changes the single members of the Markov chain
are correlated. Therefore, it will take some time to reach equilibrium and to
forget the initial condition, which usually have a dramatically different
value of the observable $L$ than the typical realizations for the biased
ensemble at the given $\Theta$. The links in the chain until equilibration have
to be discarded. Also in equilibrium subsequent samples will be correlated,
which could lead to an underestimation of the statistical error. Therefore we use the
integrated autocorrelation time \cite{newman1999monte} to only consider
statistically uncorrelated samples for our results.

This technique was used before to study diverse problems ranging from traffic models \cite{staffeldt2019rare},
over sequence alignment \cite{fieth2016score} to the ground state energy distribution
of a random energy model \cite{Schawe2018ground}. Even more general
formulations of this algorithm exist for problems, where correct change moves
are not easily defined, e.g., growth processes \cite{SHM2018,schawe2019true,hartmann2014high}.

For the results obtained here (see Figs.~\ref{fig:ld:n} and \ref{fig:ld:t}), we
used about 20 temperatures per system size, and chains
with a length of $10^5$ sweeps, each sweep being $n$ respectively
$\lceil t \rceil$ change move proposals. Equilibration was always achieved in
less than $1000$ sweeps and did not pose any problems. The autocorrelation times
range from less than 10 sweeps to around 100 for temperatures close to 0, such
that even our very far tails consist of multiple thousand independent samples.

\subsection{Mean and variance of the perimeter of the convex hull of RTP}
\label{Num_Verific}

\begin{figure}[htb]
    \centering

    \subfigure[\label{fig:cmp:n}]{
        \includegraphics[scale=1]{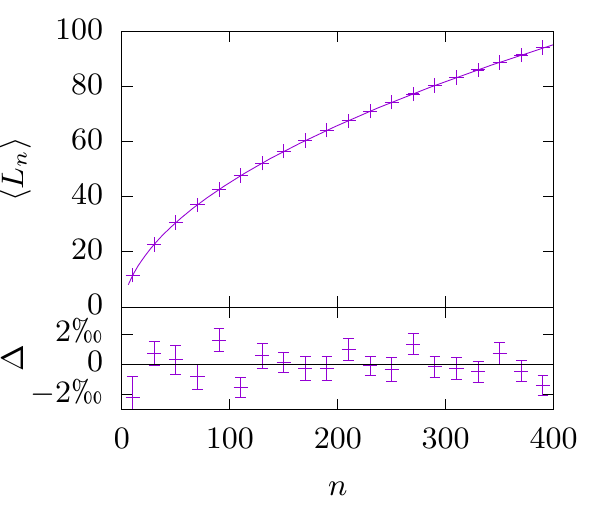}
    }
    \subfigure[\label{fig:cmp:t}]{
        \includegraphics[scale=1]{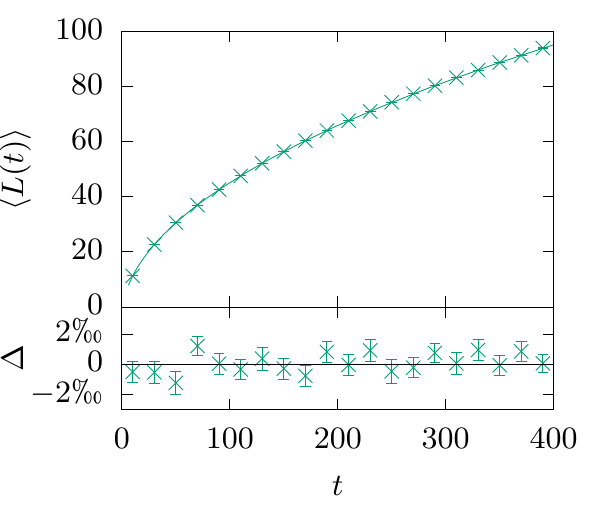}
    }

    \caption{\label{fig:cmp}
        Left: Comparison of the exact mean perimeter $\left< L_n \right>$ of
        Eq.~(\ref{fixed_n_peri}) (lines) for the fixed-$n$ ensemble with
        simulation results (symbols).
        Right: Comparison of the exact mean perimeter $\left< L(t) \right>$ of
        Eq.~(\ref{fixed_t_peri}) (lines) with simulation results (symbols).
        The errorbars are the standard errors of our simulation results
        and indicate excellent agreement, which is especially well visible when
        considering the relative difference $\Delta$ between our analytical and
        numerical result shown at the bottom of both plots. The shown data are
        for $v_0 = 1$ and $\gamma = 1$.
    }
\end{figure}

\begin{figure*}[htb]
    \centering

    \subfigure[\label{fig:mean_and_var:mean}]{
        \includegraphics[scale=1]{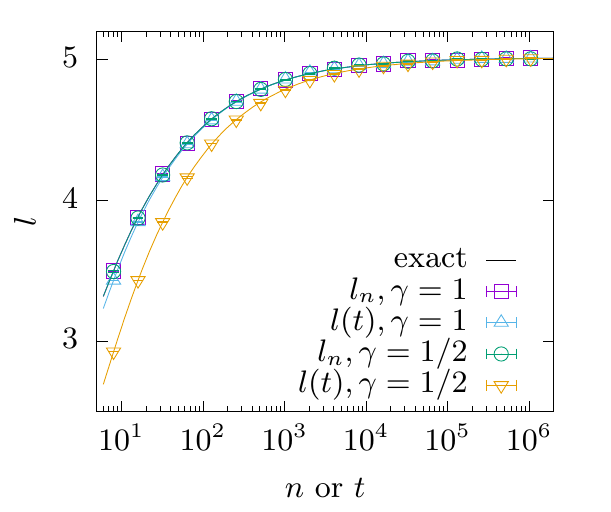}
    }
    \subfigure[\label{fig:mean_and_var:var}]{
        \includegraphics[scale=1]{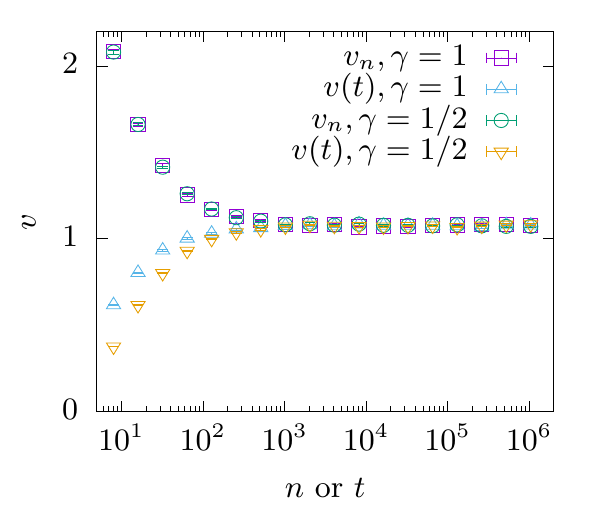}
    }

    \caption{\label{fig:mean_and_var}
        Left: Behavior of the mean perimeter of the convex hull. The exact values
        from Eqs.~\eref{fixed_n_peri} and \eref{fixed_t_peri} are shown as lines.
        Symbols show values for the scaled mean perimeter of both ensembles
        $l_n = \avg{L_n}/(\sqrt{n}\sigma_j)$ and $l(t) = \avg{L(t)}/(\sqrt{t}\sigma_j)$.
        Right: Behavior of the variance of the perimeter of the convex hull.
        Symbols show values for the scaled variance of the perimeter of both ensembles
        $v_n = \var{L_n}/(n \sigma_j^2)$ and $v(t) = \var{L(t)}/(t \sigma_j^2)$.
        Two values $\gamma = 1$ and $\gamma = 1/2$ are shown.
        For motivation of the
        unusual scaling see text. Errorbars are far smaller than the symbols.
    }
\end{figure*}

First, we compare the analytical result for the mean perimeter for
the fixed-$n$ ensemble Eq.~(\ref{fixed_n_peri}) and the fixed-$t$ ensemble
Eq.~(\ref{fixed_t_peri}) to our
simulations. These comparisons are shown in Fig.~\ref{fig:cmp}.
The agreement between numerical and analytical results
is excellent for all values of $n$, and respectively for all $t$.
Figure~\ref{fig:mean_and_var} shows that this agreement spans over
6 orders of magnitude of our simulations.

Next, we give an outlook for the mean perimeter of the convex hulls in both ensembles, as well
as the variances of the perimeter in both ensembles,  in Fig.~\ref{fig:mean_and_var}.
Here, we will scale the results in an unusual way, to enable the visualization
of different values of the parameter $\gamma$ and a very large range of sizes
in a way which enables a qualitative comparison of the different ensembles.
Since we are mainly interested in the behavior at finite sizes and less in the
asymptotic behavior, which should converge to the known case of Brownian motion,
we will remove the asymptotic growth by showing $\left< L_n \right> / \sqrt{n}$,
respectively $\left< L(t) \right> / \sqrt{t}$. Note that this way, for large
sizes the curves will converge to a limit value and therefore compress the
$y$-axis to allow the observation of fine details. For different values
of $\gamma$ these limit values are different, but connected by a simple
relation depending on the variance of the jump-length distribution $\sigma_j$,
i.e., $\sigma_j = \frac{v0}{\gamma}$ (cf.~Eq.~(\ref{varx_n})) for fixed $n$ and
$\sigma_j = \frac{v0}{\sqrt{\gamma}}$ for fixed $t$.
Therefore, we also scale with this factor which will lead to all values for the
perimeter converging to the same limit value, which makes it
possible to directly observe the qualitative influence of $\gamma$ on the
finite-size behavior. Also note the logarithmic $x$-axis to allow for a higher
visual resolution at small times.
Also, the variance is scaled appropriately with $\sigma_j^2$ and reproduces the limit values which were
also measured before \cite{Claussen2015Convex,schawe2017highdim}, but are not
known exactly even for the Brownian motion case (the only currently known
exact result for variance is for the perimeter of the convex hull of
Brownian bridges \cite{Goldman1996spectrum}.)

The finite-size behavior of the mean perimeter shown in the left
panel of Fig.~\ref{fig:mean_and_var} mirrors the result obtained in the
previous section for the perimeter: The behavior of the fixed-$t$
ensemble changes with $\gamma$ at small times, and converges to Brownian
motion for long times. The fixed-$n$ ensemble becomes indistinguishable
for all values of $\gamma$ when using this rescaling. This is consistent with
the result stated in Eq.~(\ref{fixed_n_peri}).

The behavior of the variance in the right panel shows the same effect. But interestingly, in
contrast to the mean, where the fixed-$n$ and fixed-$t$ ensemble behave similarly,
the variance approaches for the perimeter the asymptotic value from different directions
depending on whether $t$ is fixed (from below) or $n$ is fixed (from above).

We also estimated the asymptotic values $\avg{L_n} / \sqrt{n} \to \mu_\infty$
for the means and $\var{L_n} / n \to \sigma^2_\infty$ for the variances and
analogously for the fixed-$t$ ensemble. All coincide within statistical errors
with the current best estimates for Brownian motion, as expected.
This is mainly a cross check, to establish the good quality
of our data. They are obtained by fits to the form
\begin{equation}
    L(t) = \mu_\infty + a t^{-0.5} + b t^{-1},
    \label{eq:fit}
\end{equation}
and analogously for fixed $n$, as well as for $\sigma_\infty^2$. They are
shown in table \ref{tab:mean} in comparison to the current best values for them.

\begin{table}[htb]
    \caption{\label{tab:mean}
        Table of the asymptotic values in comparison to the known values for
        Brownian motion (BM). Note that the $\gamma = 1/2$ case corresponds
        for the fixed-$n$ ensemble to $\sigma_j = 2$ and for the fixed-$t$
        ensemble to $\sigma_j = \sqrt{2}$. Dividing the shown values with the
        corresponding power of $\sigma_j$, leads to values compatible with Brownian
        motion. The asymptotic values are obtained by fits to Eq.~(\ref{eq:fit}) for
        $n \ge 100$ respectively $t \ge 100$ and show goodness of fit $\chi_\mathrm{red}^2$
        between $0.5$ and $1.4$.
    }
    \centering
    \begin{tabular}{lll}
        \toprule
            &\multicolumn{2}{c}{$L$}\\
            &\multicolumn{1}{c}{$\mu_\infty$} & \multicolumn{1}{c}{$\sigma_\infty^2$} \\
        \midrule
            BM \cite{Eldan2014Volumetric,schawe2017highdim}, & 5.0132..  & 1.077(1)\\
            fixed-$n$, $\gamma = 1, \mu$   & 5.014(1)  & 1.078(2)  \\
            fixed-$t$, $\gamma = 1, \mu$   & 5.011(1)  & 1.076(2)  \\
            fixed-$n$, $\gamma = 1/2, \mu$ & 10.027(3) & 4.306(10) \\
            fixed-$t$, $\gamma = 1/2, \mu$ & 7.092(2)  & 2.152(5)  \\
        \bottomrule
    \end{tabular}
\end{table}

\subsection{Full distributions}
\label{sec:num_dist}

Here, we study the whole distribution of the perimeter of the convex hulls
of RTP. We concentrate on two regions: First, we examine the region around the mean
describing typical fluctuations, where
we show the distribution for a large range of sizes. Second,
we investigate the intermediate and far tail, which requires a much larger
numerical effort, and is therefore restricted to small sizes.

When visualizing the distribution, we will scale it with the expected perimeter.
The reasoning is similar to the unusual scaling of Fig.~\ref{fig:mean_and_var}:
First, this enables us to compare the distribution for values of $n$ and $t$ of
different magnitudes. Second, it was observed previously that the distributions
of standard random walks for different sizes collapse on a common curve when
scaled with the mean perimeter in the limit of Brownian motion \cite{Claussen2015Convex,schawe2017highdim}.

\begin{figure}[htb]
    \centering
    \subfigure[\label{fig:dist:n}]{
        \includegraphics[scale=1]{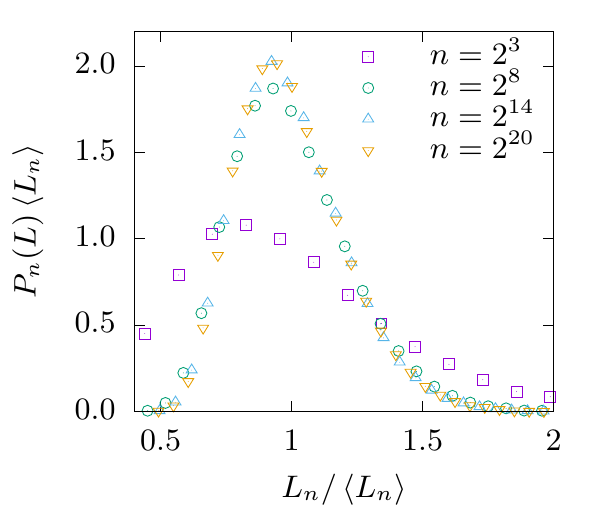}
    }
    \subfigure[\label{fig:dist:t}]{
        \includegraphics[scale=1]{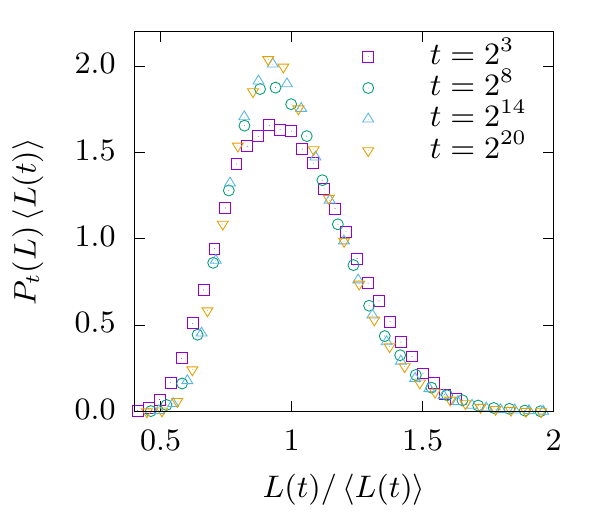}
    }

    \caption{\label{fig:dist}
        Distribution of the perimeter $L$ of the RTP with $v_0 = 1$ and $\gamma = 1$.
        Both fixed-$n$ and fixed-$t$ are shown in comparison and scaled with
        the leading order of their mean. Note that a scaling with their exact
        mean does also not lead to a collapse of the small sizes (not shown).
    }
\end{figure}

Unsurprisingly the distributions for the perimeter of
the RTP's convex hull to collapse in the limit of large sizes, which corresponds
to Brownian motion, on the same curve when rescaled with the exact mean perimeter
according to Eqs.~\eref{fixed_n_peri} and \eref{fixed_t_peri}. This is shown in Fig.~\ref{fig:dist}.
While small sizes show significant deviations from this limit shape, it is
interesting that the fixed-$t$ ensemble is apparently converging much faster.

\begin{figure*}[htb]
    \centering
    \subfigure[\label{fig:ld:n}]{
        \includegraphics[scale=1]{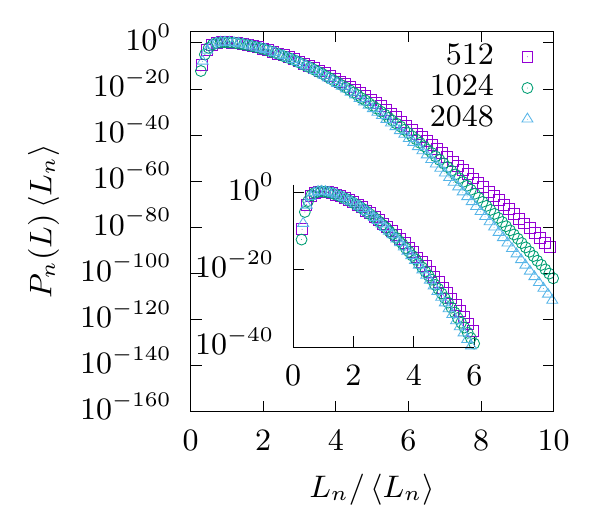}
    }
    \subfigure[\label{fig:ld:t}]{
        \includegraphics[scale=1]{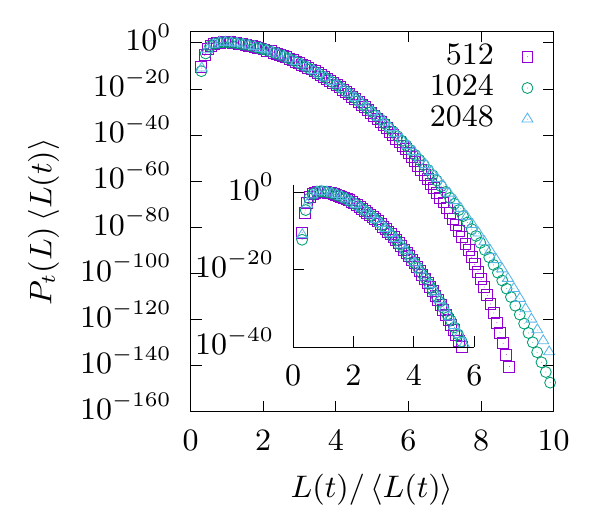}
    }

    \caption{\label{fig:ld}
        Probability density function of the perimeter of the convex hull around
        RTP including high quality statistics for extremely rare configurations
        with very large and very small hulls.
        Left for the fixed-$n$ ensemble, right for the fixed-$t$ ensemble.
    }
\end{figure*}

With the large-deviation approach, we are able to sample the distribution of the
perimeter over more than one hundred decades, see Figs.~\ref{fig:ld:n} and \ref{fig:ld:t}.
Examining the tails of the distribution, one sees that the scaling of the
distribution with its mean value works not only in the high probability region,
but also in the intermediate tails, as shown in the insets.
In the far tails this collapse ceases to work.
Also in the far tail, we can observe a curious difference
between the fixed-$t$ and fixed-$n$ ensemble. For fixed-$n$ smaller numbers of
tumblings lead to a larger value of the scaled probability for extremely large hulls, while
for fixed-$t$ a longer total time leads to a lower value of the scaled probability for
extremely large hulls.

As expected due to previous results on convex hulls of random walks, which
converge to Brownian motion in the long time limit, we
test whether a \emph{large deviation principle} holds \cite{Touchette2009large}. That is,
the \emph{rate function} $\Phi$, defined by
$P_n(L) \approx \exp{\left( -n \Phi(L_n) \right)}$, respectively
$P_t(L) \approx \exp{\left( -t \Phi(L(t)) \right)}$ (without explicit dependency on $n$ or $t$)
for large $n$ respectively $t$, exists in the right tail. As an intuitive
interpretation, the rate function determines how fast the probability density
decays in the tails with increasing $n$ respectively $t$. For the present
distribution we can verify this directly
using the tails of the distributions we obtained: The \emph{empirical rate functions}
$\Phi_n(L/n)$, respectively $\Phi_t(L/t)$ calculated from our data (not shown here)
collapse on a power law with exponent $2$ over a large range
(including the full right tail) where they are independent of the
system size, which is a strong hint that these forms are the actual $n$, respectively $t$
independent rate functions. The exponent $2$ is also compatible with previous
results \cite{Claussen2015Convex,schawe2017highdim}.

\begin{figure*}[htb]
    \centering
    \subfigure[\label{fig:corr:n}]{
        \includegraphics[scale=1]{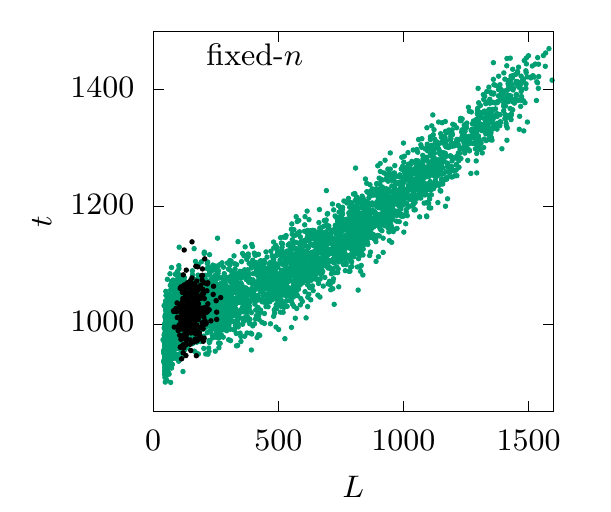}
    }
    \subfigure[\label{fig:corr:t}]{
        \includegraphics[scale=1]{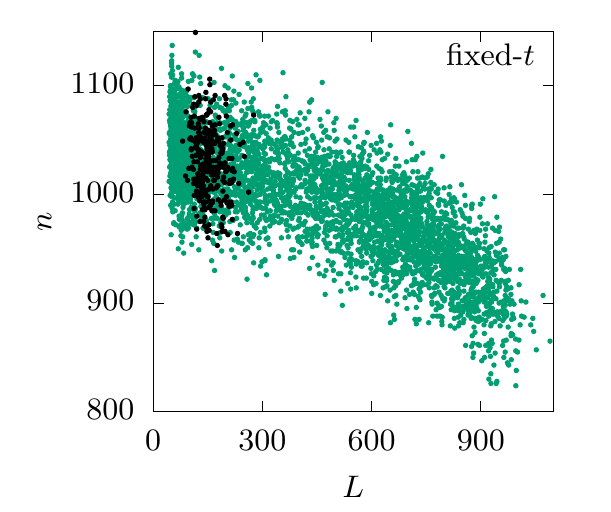}
    }

    \caption{\label{fig:corr}
        Left: Strong correlation of the total time $t$ with the perimeter for the
        fixed-$n$ ensemble. Right: Anticorrelation of the number of tumble events
        $n$ with the perimeter for the fixed-$t$ ensemble.
        Data for $\gamma=1$ und $n=1024$, respectively $t=1024$.
        Events with a high probability (gathered using simple sampling) are
        marked black.
        Note that the density of points does not mirror the actual probability
        of the events.
    }
\end{figure*}

To understand what characterizes the instances of extremely large and
extremely small
convex hulls, we measured during the large-deviation Markov chain MC simulations
for the fixed-$n$ ensemble the total time
traveled $t$ and for the fixed-$t$ ensemble
the number of tumble events $n$, for each realization encountered, respectively.
To end up with a very large hull, the RTP can either make very large steps,
which become exponentially rare with their length, or can take steps
persistently into one direction, which also becomes exponentially improbable
with the number of tumble events. If we look in Fig.~\ref{fig:corr} at the
correlations between the perimeter of the hull and the number of tumble
events $n$ (for fixed-$t$), respectively the total time of the
walk $t$ (for fixed-$n$), we can identify the mechanism in both ensembles
leading to extremely large hulls. In the fixed-$n$ ensemble the perimeter of
the hull is strongly correlated with the total time, i.e., the size of the hull
is inflated by taking longer steps. The question is whether this correlation
is sufficient to explain the large-deviation behavior.
For a configuration of fixed ``shape'' (i.e. directions), if one blows
up all traveling times by a certain factor, this would result in
a growth of the perimeter by the same factor. Now, the extreme configurations
exhibit about a ten times higher perimeter, but only about a factor
of 1.4 longer time.
Hence, rare configurations are not only
characterized by atypical large times between two tumble events, but also
by rare combination of chosen directions.

Correspondingly, in the fixed-$t$ ensemble, the number of
steps is anticorrelated with the perimeter. A smaller number of tumble
events, while having the same total traveling time, leads intuitively to
more extended RTP trajectories, i.e., to larger perimeters.
Nevertheless, since the extreme perimeter values are
much larger than the typical ones and the number of tumble events
is only moderately decreased,
the large-deviation behavior is also for this ensemble influenced
by rare combination of chosen directions.

\section{Conclusion}\label{sec:conclusion}
In this paper we have studied the statistics of the perimeter of the
convex hull of a single run-and-tumble particle in two dimensions. This run-and-tumble
particle moves with constant speed (ballistically) during an exponentially distributed
run time and
changes its direction of motion at tumble events instantaneously by choosing
a new direction at random. We derived exact
expressions for the mean perimeter of the convex hull of its trajectory for two ensembles,
with a fixed number of tumblings and with a fixed total time.
Our numerical simulations in both ensembles are in excellent agreement with analytical
results. For higher moments of the perimeter we could not derive exact results, but
we obtained numerical results with very high precision. In particular, for both ensembles
we presented detailed numerical studies of the variances as well as the full
probability distributions of the perimeter of the convex hull. Deriving analytically
the higher moments as well as the full distribution of the perimeter remains a challenging
open problem. In particular, it would be interesting to
understand in more detail what leads to the smaller variance
of the perimeter of the fixed-$t$ ensemble in comparison to the fixed-$n$
ensemble for finite sizes.

Another closely related observable is the area of the convex hull of the trajectory.
For Brownian motion and other stochastic processes, the mean area of the associated
convex hull has been calculated analytically (for a review see~\cite{ch2}). In fact,
a second formula due to Cauchy for the area of an arbitrary
convex curves in $2d$ was applied to convex hulls~\cite{ch1,ch2} that allows one
to compute the mean area of the convex hull of an arbitrary stochastic process in $2d$.
Let $\{X_i,Y_i\}$ (with $i=1,2,\ldots,n$) denote the $n$ vertices of a stochastic process in $2d$
where $i$ labels the time. Let $\cal C$ denote the convex hull of these vertices. Then the
mean area of the convex hull of these vertices is given by~\cite{ch1,ch2}
\begin{equation}
    \langle A_n\rangle = \pi\, \left[ \langle M_n^2\rangle - \langle Y_{m}^2 \rangle\right]\, .
    \label{area.1}
\end{equation}
where $M_n= {\rm max}\, \left[X_1,X_2,\ldots,X_n\right]$ denotes the maximum of the $x$-coordinates which
is achieved at step $m$. In formula (\ref{area.1}) $Y_m$ denotes the $y$-coordinate at step $m$, i.e,
at the time when the $x$-coordinate achieves its maximum.
This formula (\ref{area.1}) is valid for any set of random points in $2d$~\cite{ch2}, and hence
it is also valid for the $2d$ RTP process. However this analytic computation of the mean area is
not that easy and we have not done it yet.
However, we could easily obtain the
mean area, as well as its higher moments and the full distribution numerically, using the same method that
we used for the perimeter (not presented in this paper).
Hence, it would be interesting if this mean area can be computed analytically to
compare with our simulation data.

Finally, in this paper, we have considered the simplest version of the run-and-tumble model
where the walker tumbles instantaneously. In more realistic models, there is an additional
waiting time after each run during which the particle tumbles, before starting a new run.
It would be interesting to see how the mean perimeter of the convex hull gets affected by the finite waiting time.
Evidently, the result for the fixed-$n$ ensemble will not depend on the waiting time. However, in the fixed-$t$
ensemble, the mean perimeter will certainly depend on the waiting time, and it would
be interesting to compute this.

\ack
HS and AKH thank the German Science Foundation (DFG) for financial support through
the grant HA 3169/8-1.
The simulations were performed at the HPC cluster CARL, located at the
University of Oldenburg (Germany) and funded by the DFG
through its Major Research Instrumentation Programme
(INST 184/108-1 FUGG and INST 184/157-1 FUGG) and the Ministry of
Science and Culture (MWK) of the Lower Saxony State.

\section*{References}
    \bibliographystyle{iopart-num}
\providecommand{\newblock}{}

\end{document}